\documentclass[11pt,twoside]{article}

\usepackage{asp2006}
\usepackage{epsf}
\usepackage{lscape}

\begin{document}
\title{Star Formation History across the Magellanic Clouds and other
  Local Group galaxies}   
\author{M.-R.L. Cioni}   
\affil{SUPA, School of Physics, University of Edinburgh, Institute for
  Astronomy, Blackford Hill, Edinburgh EH9 3HJ, United Kingdom}    

\begin{abstract} 
  The $K_s$ magnitude distribution of asymptotic giant branch stars
  provides important constraints on the spatial variation of
  metallicity and mean-age of the stellar population of galaxies.
  Here, I present results of the investigation of these parameters
  across the Magellanic Clouds, M33, NGC 6822 and SagDIG and I discuss
  the tremendous improvement that new wide-field near-infrared instruments
  (i.e.~VISTA) will provide on our understanding of the global
  intrinsic as well as dynamical history of these systems.
\end{abstract}



\section*{Introduction}

The Magellanic Clouds (MCs), like other galaxies in the Local Group
(LG), represent the best place to study properties of individual stars
inhabiting them and from these deduce properties of the galaxies as a
whole. The main advantages are a nearby distance of the systems and a
relatively low extinction along the line of sight.  This research aims
to obtain a global picture of the history of formation and evolution
of galaxies which requires to understand the distribution of age,
chemical abundance and kinematics of both stars and gas.  This
contribution highlights the importance of asymptotic giant branch
(AGB) stars to make a step forward in the understanding of galaxy
formation and evolution.  In particular, towards obtaining a global
picture of galaxies of the LG.

\section*{Learning from AGB stars}
There are different galaxy properties that can be inferred from the
study of AGB stars, among these: their number density constrains the
galaxy structure; the ratio between carbon-rich (C-rich) and
oxygen-rich (O-rich) AGB stars, the C/M ratio, constrains the iron
abundance; the $K_s$ magnitude distribution constrains the star
formation history of the galaxy, i.e.~using up-to-date theoretical
models allows us to derive the mean-age and the metallicity ([Fe/H])
of the underlying stellar population (not only of evolved giants) and
with the availability of low- an high-resolution spectra we can
constrain their kinematic and chemistry.  The study of how the above
quantities vary across the parent galaxy provides crucial information
about population gradients.

The first step to use AGB stars as probes of galaxy evolution is to
select them among other types of stars.  AGB stars are easily
distinguished from stars in other phases of evolution using
colour-magnitude diagrams (CMDs). For example in the ($I-J$, $I$)
diagram they occupy a plume above the tip of the red giant branch
(RGB) extending to red colours \citep*{ciha00}. The near-infrared CMD
($J-K_s$, $K_s$) provides the additional statistical distinction
between C-rich and O-rich AGB stars (\citeauthor{cial06a} 2006a).
C-rich stars are redder and somewhat brighter than O-rich stars. These
can also be distinguished using other criteria (see Groenewegen, {\it
  this proceeding}).

The number density of AGB stars across a galaxy often describes a
smooth surface distribution similar to the one obtained from, on
average older, RGB stars and contrary to the clumpier and irregular
distribution traced by younger stars (\citeauthor*{ciha00} 2000;
\citeauthor{ciha05} 2005).  Distance indicators such as the mean
magnitude of C-rich stars in a narrow range of ($J-K_s$) colours or
the mean magnitude of AGB stars in a narrow range of ($I-J$) colours
as well as the tip of the RGB have been used to probe the inclination
and position angle of the Large Magellanic Cloud (LMC;
\citeauthor{vdmci01} 2001) and to correct for
differential reddening across NGC 6822 (Cioni et al.~{\it in
  preparation}).

The C/M ratio is an important indicator of the distribution of iron
across a galaxy. A calibration of this ratio versus the [Fe/H]
abundance shows a spread of $~0.75$ dex within each MC and a larger
spread across NGC 6822.  In particular, the metallicity across the LMC
decreases radially \citep{ciha03}.  This result was later confirmed by
\citet{al04} using RGB stars.  A preliminary study of the distribution
of the C/M ratio across the nearby spiral M33 shows metal poor regions
tracing the major spiral arms of the galaxy and a $\Delta$[Fe/H]$=0.6$
dex (Cioni et al.~{\it in preparation}).  This pattern
supports the findings of \citet{ro05} of an increasing C/M ratio up to
$20^{\prime}$ from the centre and a further flattening induced by the
flattening of the rotation curve which causes gas mixing (i.e.~metal
enrichment) at larger distances from the centre.

\section*{The $K_s$-method}

The observed $K_s$ magnitude distribution of AGB stars interpreted
using theoretical distributions produced from stellar evolutionary
models (\citeauthor{gi05} 2005, \citeauthor{ma03} 2003) allows us to
measure both the metallicity and the mean-age of the underlying
stellar population. This $K_s$-method has allowed to obtain for the
first time a global description of the distribution of these
parameters across the MCs (\citeauthor{cial06a} 2006a, 2006b).

\begin{figure}[!ht]
\plotone{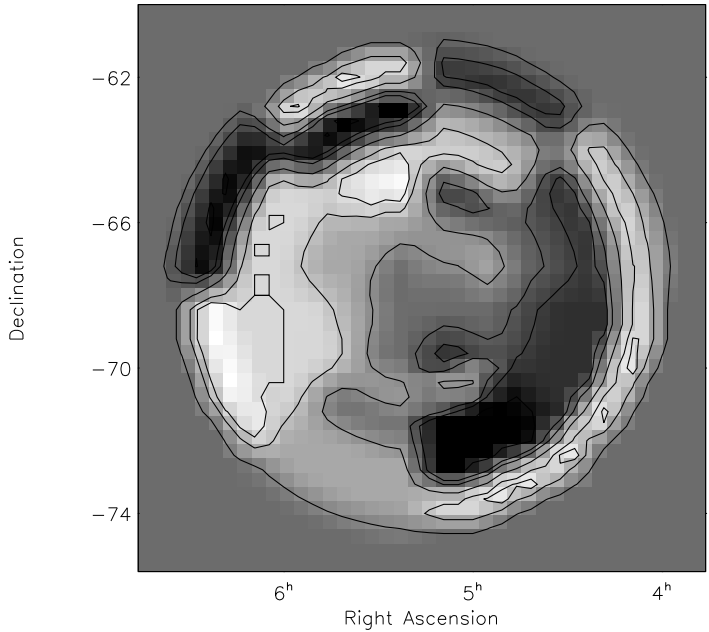}
\caption{Distribution of the most probably mean-age of the stellar
  population across the LMC.  Contours are at: $2$, $3$, $5$, $7$,
  $8$ and $10$ Gyr. These map has been obtained by comparing the
  observed $K_s$ magnitude distribution of C-rich AGB stars with
  theoretical distributions and it is corrected for the sky
  orientation of the LMC. Darker regions correspond to higher
  numbers. This figure is published in \citeauthor{cial06c} (2006c).}
\end{figure}

Results for the LMC ({\bf Fig.~1}; \citeauthor{cial06a} 2006a)
indicate a stellar population which is younger in the E and in the N
and S regions of the inner disk compared to an older population
present in the S-W of it as well as in the N of the outer disk.  The
bar region has a composite stellar population. The distribution of
metals, in agreement with the distribution suggested by the C/M ratio,
shows two central regions with a low metal content compared to a
ring-like inner disk region of higher metallicity. The metallicity is
perhaps higher in the E than in the W along this structure.  For the
SMC (\citeauthor{cial06b} 2006b), a ring-like structure, which appears
more metal rich than a S-central region, is also detected. Within this
ring the metallicity increases from S-E to N-W from a young to old
stellar population.  This effect is perhaps related to the dynamical
history of the galaxy or to its structure (Bekki \& Cioni {\it in
preparation}).

\begin{figure}[!ht]
\plottwo{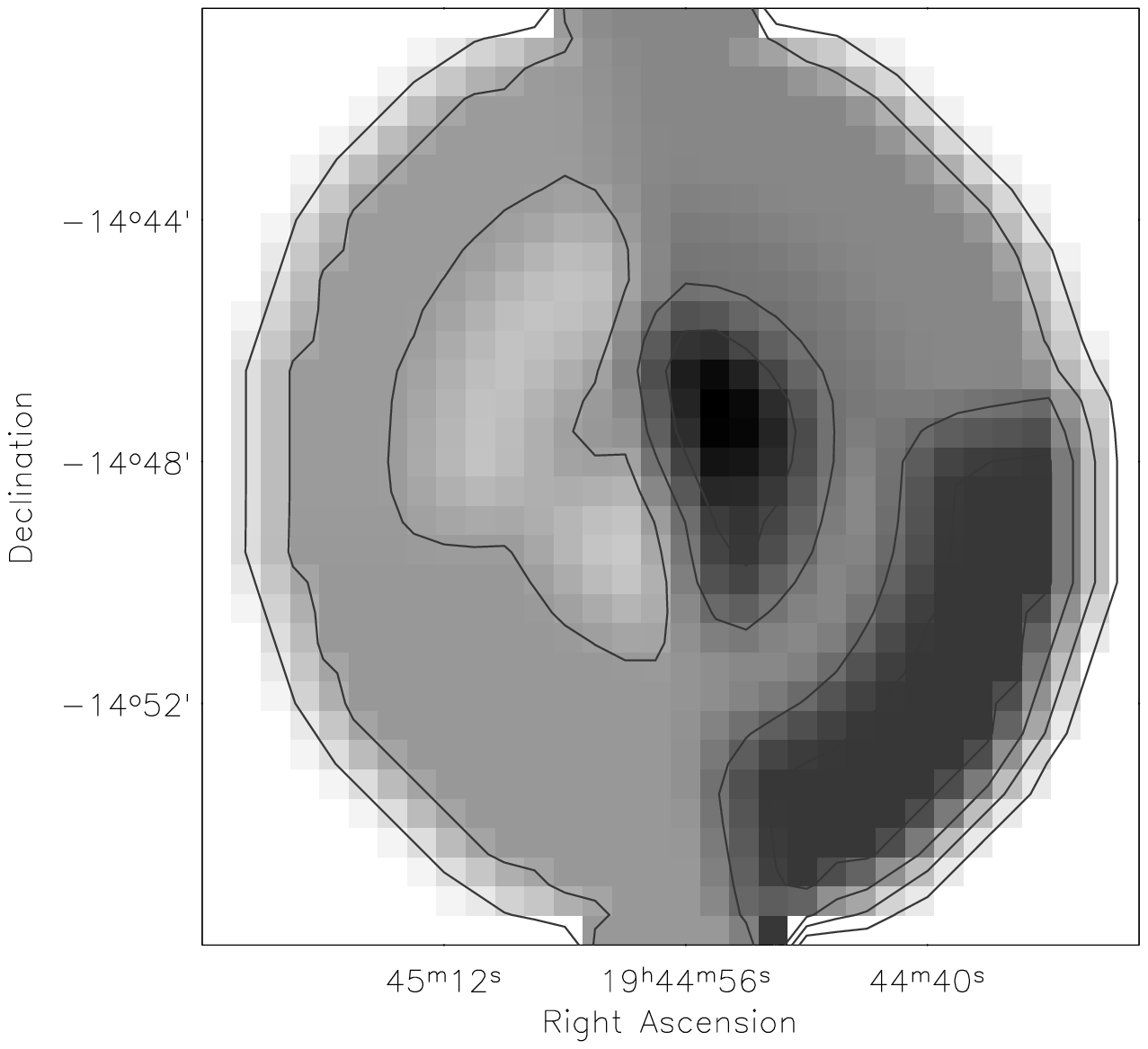}{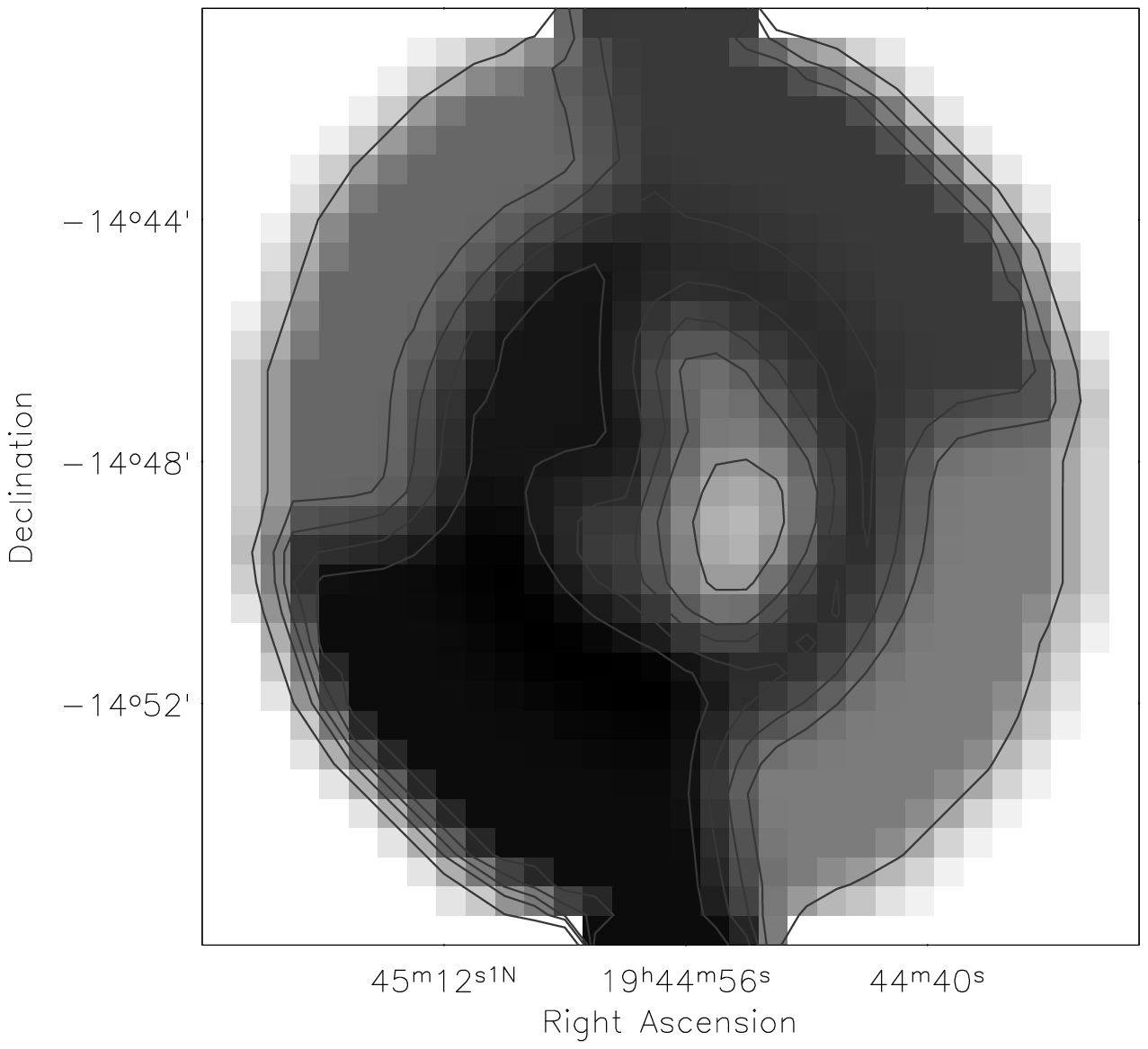}
\caption{Distribution of metallicity and mean-age across NGC
  6822. These preliminary maps are the average of an independent
  comparison between the observed $K_s$ magnitude distributions of
  C-rich and O-rich AGB stars with theoretical distributions within
  elliptical rings. They are also corrected for differential
  reddening. Metallicity contours expressed in terms of Z are from
  $0.002$ to $0.008$ with a step of $0.002$ while contours for
  mean-ages are at $4.5$, $6.5$, $7.5$, $8.5$ and $9.5$ Gyr (Cioni et
  al.~{\it in preparation}).}
\end{figure}

Recent results across another Local Group irregular galaxy of
Magellanic type, NGC 6822 ({\bf Fig.~2}; Cioni et al. {\it in
  preparation}), show an average stellar population which is about
$8.5$ Gyr old and metal poor ([Fe/H]$=-1.0$ dex). The bar appears at
least $2.5$ Gyr younger and more metal rich ([Fe/H]$=-0.6$ dex) than
the region immediately around it.  These results already account for
the differential reddening present in the galaxy \citep{bade05}.

\begin{figure}[!ht]
\plotone{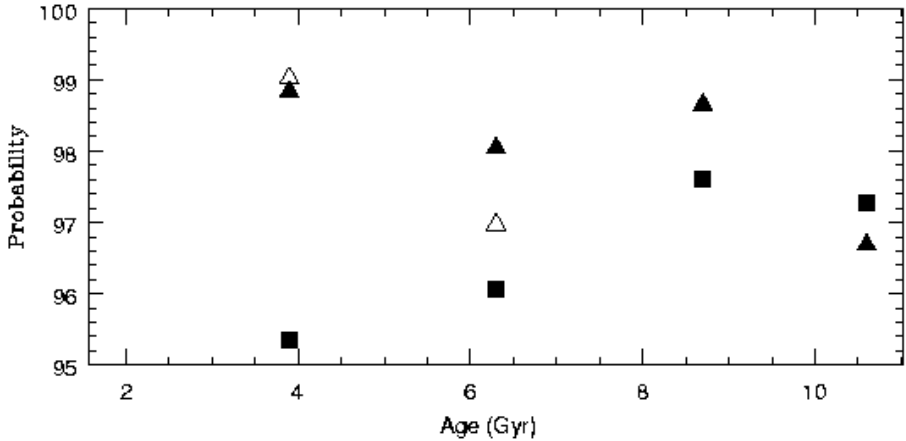}
\caption{The probability as a function of mean-age for a given
metallicity representing the stellar population of SagDIG. Different
symbols are as follows: Z$=0.0005$
(filled triangles), Z$=0.001$ (empty triangles), Z$=0.004$ (filled
squares). Other models explored in this study give a probability much
lower than the values plotted here (Gulieuszik et al.~{\it in preparation}).}
\end{figure}

The application of the $K_s$-method to the nearby spiral M33 shows a
stellar population which, in agreement with previous determinations, is
on average $7-8$ Gyr old, prior correcting for differential reddening.
This work, currently in progress, seems to indicate that the NE of the
galaxy is more obscured than the SW of it. This finding is presently
being checked against distance measurements from the literature and
may explain the cause of their disagreement.

Galaxies that do not have a numerous intermediate-age component, like
SagDIG, still produce reasonable results when the $K_s$-method is
applied to a confirmed, and more or less complete, sample of C-rich
stars ({\bf Fig.~3}; Gulieuszik et al. {\it in preparation}). In fact,
we obtain a stellar population on average young ($4$ Gyr old) and
metal poor ([Fe/H]$\le-1.3$ dex).

Summarising, the $K_s$ magnitude distribution of AGB stars allows us
to estimate variations in mean-age and metallicity across galaxies.
Modest sample of stars produce satisfactorily results. This technique
will be applied to other galaxies of the LG, as soon as suitable data
will become available, but also to more distant systems resolved into
stars, especially those systems where only AGB stars are these
resolved stars. The study of the stellar population of different
galaxies is important to understand the effect of the environment on
the formation and evolution of galaxies. Modern near-infrared
instruments resolve AGB stars up to about 5 Mpc while to reach them in
the Virgo and Fornax clusters galaxies we need an extremely large telescope.
Further, a calibration of the $K_s$-method versus the integrated
near-infrared light of stellar populations may allow us to study also
galaxies where individual stars cannot be resolved but for which
population gradients can be accessed using integral field units.

\section*{A pre-selected VISTA Public Survey proposal: the VISTA survey of the 
Magellanic Clouds (VMC; PI = Cioni + 19 Co-Is)}

Although the $K_s$-method provides information about population
gradients it is not sufficient to determine absolute ages and
metallicities. Moreover, in oder to obtain a global picture of the star
formation history we need to account for the three-dimensional
geometry of galaxies as well as for their interaction with
neighbouring systems. The determination of these parameters with an
unprecedented accuracy across the MCs is the main goal of the VMC survey. 

VISTA is a new wide-field near-infrared telescope soon to be
commissioned at the European Southern Observatory.  The VMC survey
aims to collect $YJK_s$ broad-band photometry across the Magellanic
system covering entirely the MCs, the Bridge and some regions of the
Stream. This is a deep survey that will detect sources at $K_s=20.3$
with a S/N$=10$ which is $\sim6$ magnitude fainter than the 2MASS
limit. It will take about 5 years to cover the proposed area and
multi-epoch data, especially in the $K_s$ band, will allow us to
determine the mean magnitude of short period variables (RR Lyrae and
Cepheid stars) which will be used, among other indicators, to trace
the three-dimensional structure of the system. Simulations of the
system as a whole combined with the distribution of different density
indicators (i.e. stars of a different type and nature) will allow us
to account for the effect of interactions between the MCs and the
Milky Way.

The VMC survey, of strong legacy value, represents a unique
counterpart for optical data with the same sensitivity and also for
exploiting Spitzer data.\\

VISTA in the Southern and UKIRT in the Northern hemisphere are truly
powerful instruments for the study of LG galaxies. They allow us to
resolve individual stars even in optically crowded regions and because
of the wide-field we can efficiently probe substructures (streams,
tidal tails and undiscovered bound entities) witnessing the history
of our group.

\acknowledgements 
I would like to warmly thank all co-authors of the
papers mentioned in this contribution for their active collaboration.



\begin{thebibliography}{}
\bibitem[Alves(2004)]{al04}
Alves, D.R., 2004, New Astronomy Reviews, 48, 659

\bibitem[Battinelli \& Demers(2005)]{bade05}
Battinelli, P. \& Demers, S., 434, 657

\bibitem[Cioni, Habing \& Israel(2000)]{ciha00}
Cioni, M.-R.L., Habing, H.J., \& Israel, F.P., 2000, A\&A, 358, L9

\bibitem[Cioni \& Habing(2003)]{ciha03}
Cioni, M.-R.L., \& Habing, H.J., 2003, A\&A, 402, 133

\bibitem[Cioni \& Habing(2005)]{ciha05}
Cioni, M.-R.L., \& Habing, H.J., 2005, A\&A, 442, 165

\bibitem[Cioni et al.(2006a)]{cial06a}
Cioni, M.-R.L., Girardi, L., Marigo, P., \& Habing, H.J., 2006a, A\&A, 448, 77 

\bibitem[Cioni et al.(2006b)]{cial06b}
Cioni, M.-R.L., Girardi, L., Marigo, P., \& Habing, H.J., 2006b, A\&A, 452, 195

\bibitem[Cioni et al.(2006c)]{cial06c}
Cioni, M.-R.L., Girardi, L., Marigo, P., \& Habing, H.J., 2006c, A\&A,
456, 967

\bibitem[Girardi et al.(2005)]{gi05}
Girardi, L., Groenewegen, M.A.T., Hatziminaoglou, E., \& da Costa, L., 2005, A\&A, 436, 895

\bibitem[Marigo et al.(2003)]{ma03}
Marigo, P., Girardi, L., \& Chiosi, C., 2003, A\&A, 403, 225

\bibitem[Rowe et al.(2005)]{ro05}
Rowe, J.F., Richer, H.B., Brewer, J.P., \& Crabtree, D.R., 2005, AJ, 129, 729 

\bibitem[van der Marel \& Cioni(2001)]{vdmci01}
van der Marel, R.P., \& Cioni, M.-R.L., 2001, AJ, 122, 1807
\end{thebibliography}
\end{document}